\documentclass[aps,apl,superscriptaddress,showpacs, twocolumn]{revtex4}

\usepackage{makeidx}
\usepackage{epsfig}
\usepackage{amsfonts}
\usepackage{amsmath}
\usepackage{amssymb}
\usepackage{graphicx}
\usepackage{epstopdf}

\begin{document}

\title{ Swing switching of spin-torque  valves }

\author{Tom Dunn$^1$, and  Alex Kamenev$^{1,2}$ \\
 $^1$Department of Physics, University of Minnesota, Minneapolis, Minnesota 55455, USA. \\
 $^2$Fine Theoretical Physics Institute, University of Minnesota, Minneapolis, Minnesota 55455, USA.}

\begin{abstract}
We propose a  method for inducing magnetization reversal using an AC spin current polarized perpendicular to the equilibrium magnetization
of the free magnetic layer. We show that the critical AC spin current is significantly  smaller than the corresponding DC one. The effect is understood as a consequence of the underdamped nature of the spin-torque oscillators.
It allows to use the kinetic inertia to overcome the residual energy barrier, rather than suppressing the latter by a large spin current.
The effect is similar to a swing which may be set into high amplitude motion by a weak near-resonant push.
The optimal AC frequency is identified as the upper bifurcation frequency of the corresponding driven nonlinear oscillator.
Together with fast switching times it makes the perpendicular AC method to be the most efficient way to realize spin-torque memory valve.
\end{abstract}

\maketitle

\section{Introduction}
\label{sec-intro}

Spin-torque (ST) induced switching in magnetic tunnel junctions (MTJ) has been a subject of intense experimental
and theoretical research since its inception by  Slonczewski \cite{Slonczewski96} and
Berger \cite{Berger96} over a decade ago. It has been demonstrated \cite{Myers99, Katine00,Hosomi05, Matsunaga09} to be a viable candidate
for fast, scalable and non-volatile memory storage.
Early  MTJ devices \cite{Grollier01,Ozyilmaz03,Urazhdin03,Seki06, Yuasa07} utilized two ferromagnetic layers, separated by a non-magnetic layer, with magnetization axes aligned
parallel to each other and orthogonal to the vertical axis of the pillar, Fig.~\ref{fig:MTJ}a (hereafter referred to as the parallel method).
These devices tend to have switching times longer than $1ns$ \cite{Sun00, Lee09, Nikinov10} as well as a broad switching time distribution \cite{Myers02}.
These characteristics come from an initial incubation time during the switching process when thermal noise
provides for an initial incipient misalignment between the two layers \cite{Li04, Apalkov05}. Sub nanosecond
switching times are possible \cite{Bedau10}, but require rather strong currents and are far from the optimal efficiency \cite{Dunn11}.
It was recently proposed by Kent {\it et all} \cite{Kent04}  that
adding a third ferromagnetic layer, with an easy axis perpendicular to the other two, i.e. along the vertical axis of the MTJ, Fig.~\ref{fig:MTJ}b, may reduce switching times below $1ns$ \cite{Lee09, Nikinov10, Liu10, Rowland11}.  The improvement is due to the fact that the torque acts perpendicular to the initial magnetization direction and therefore does not rely on thermal noise to initiate the switch.
This approach, here known as the DC perpendicular method, allows magnetic reversals of the free layer at a fraction of the energy cost, i.e. associated Joule heating \cite{Rowland11}.
While a significant improvement over the parallel method, it has it's own
deficiency. The reason is that the magnetization tends to precess around the easy axis direction. As a result the perpendicular torque
acts in the wrong direction for a half of the precession period \cite{Kent04}.
Therefore the protocol  is most efficient if it can accomplish the switch
during the half of the precession period. This dictates rather high values of the required spin current.

\begin{figure}[h]
  \begin{centering}
  \includegraphics[width=9cm]{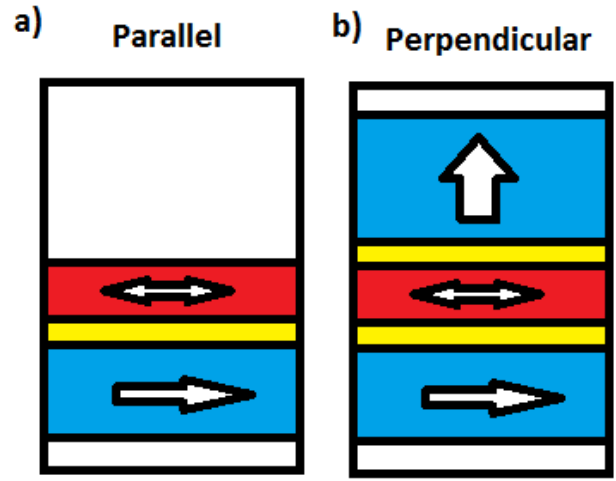}
  \par\end{centering}
  \caption{(Color online) (a) Diagram of parallel MTJ. (b) Diagram of perpendicular MTJ. Blue and red regions represent fixed and free layers respectively. Yellow regions are non-magnetic spacers. Current flows along the vertical axis of the pillars.}\label{fig:MTJ}
  \label{fig:MTJ}
\end{figure}

In this paper we suggest an alternative approach to magnetic reversal in
MTJs using an AC current in the perpendicular configuration
(here known as the AC perpendicular method) with the driving
frequency near the natural precession frequency of the free layer.
The use of microwave frequency AC spin-current to influence the magnetization
dynamics of the free layer has been studied in a number of works in recent years.
References \cite{Sankey06,Mourachkine08} employed a weak AC signal
to excite ferromagnetic resonance excitations of the free layer, detected
through the induced DC voltage in the perpendicular geometry.  References
\cite{Tulapurkar05,Georges09} performed similar experiments in the parallel geometry.
References \cite{Florez08,Cheng10} showed that in the parallel geometry an
AC signal on top of a DC pulse may facilitate stochastic thermal switching events between the
two metastable states. And reference \cite{Cui08} showed
that an AC signal prior to a DC pulse can also facilitate switching by reducing the
critical current in the parallel geometry.

Here we show that a properly tuned purely AC current in the perpendicular
geometry may serve as the most energy efficient way to switch ST memory
valves.
Our simulations show that this method produces fast switching times, below $1ns$, and
allows for switching currents about a {\em factor of two below} the critical current
of the equivalent DC device. As a result it accomplishes the switch with a significantly
smaller energy dissipation than the DC method.
We develop a theory which   explains the behavior observed in the simulations
and serves as a useful guide to optimize the system parameters.
It is based on the fact that the energy  of the Stoner-Wohlfarth (SW) orbital and the
relative phase between the magnetization and the external AC drive form a  canonical  action-angle pair \cite{Dunn12}.
As a result,
the system may be described as a weakly damped non-linear oscillator.
Its effective potential landscape is affected by both the amplitude and the frequency of the AC spin current.
The key observation, explaining the critical current reduction, is the inertial, i.e. underdamped, nature
of this effective oscillator. It allows the switching to occur without fully suppressing the
energy barrier, but rather overcoming it by inertia. The effect is similar to setting a swing into a rotation by
weak near-resonance AC push.

\section{Simulations of the perpendicular spin-torque valve}

\begin{figure}[h]
  \begin{centering}
  \includegraphics[width=9cm]{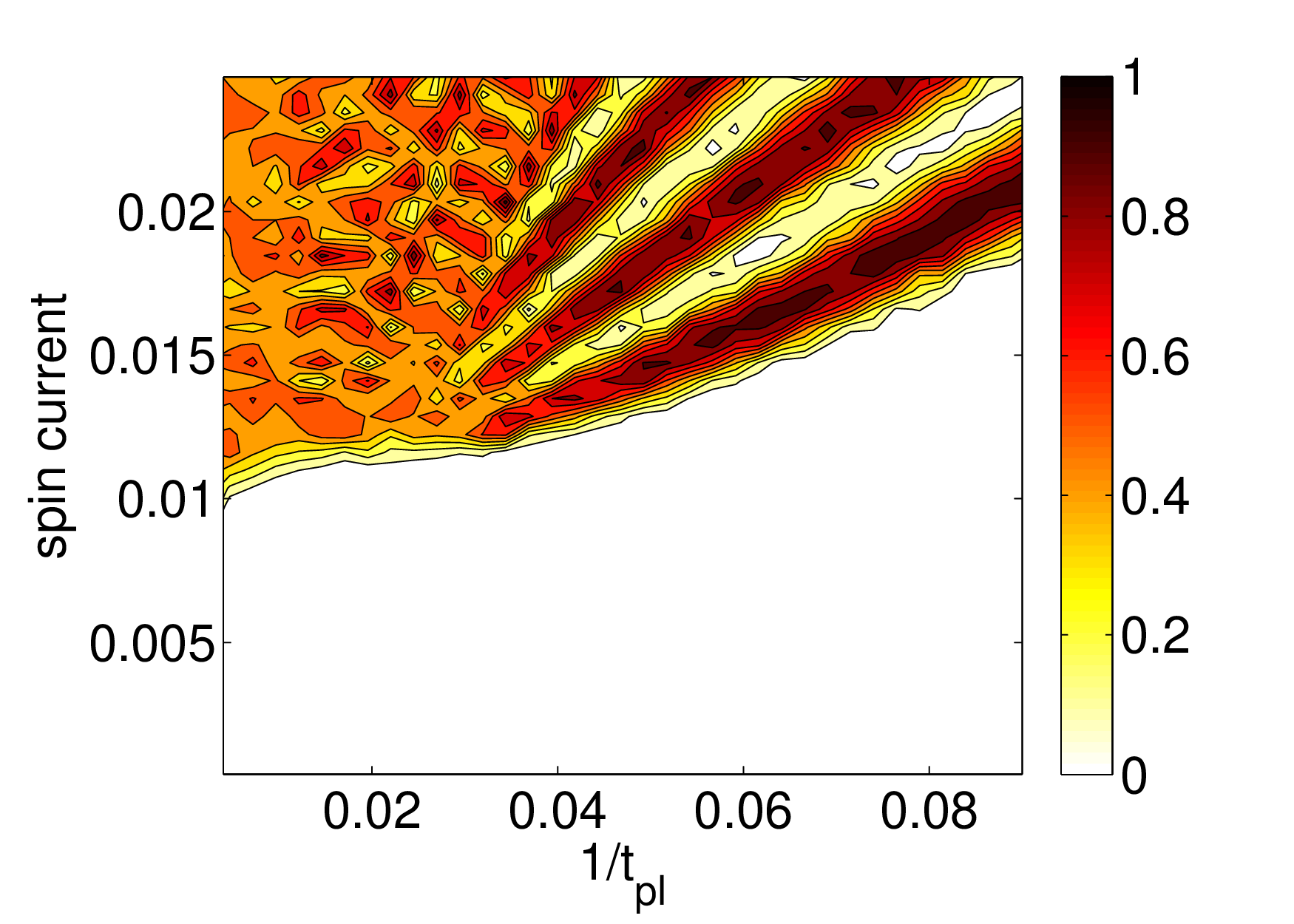}
  \par\end{centering}
  \caption{(Color online) Switching probability of DC perpendicular valve as a function of spin current amplitude ($\mathcal{I}_s/M_s$) and an inverse pulse duration $t_\mathrm{pl}$ (in units of $\gamma M_s$) with $H_x = M_s$, $H_z = 0.033 M_s$, $\alpha = 0.015$, and $T = 300K$.}\label{fig:DC_probability}
\end{figure}

To model the magnetic switching we treat the free layer as a single magnetic
domain with a constant saturation magnetization $M_s$ and  magnetization direction specified by a time-dependent unit vector $\mathbf{\hat{m}}(t)$. Its dynamics is described by the Landau-Lifshitz-Gilbert equation with Slonczewski
spin torque term
\begin{eqnarray}
\label{m-det}
\frac{d\mathbf{\hat{m}}}{dt} = \mathbf{\Gamma}_\mathrm{LL} + \mathbf{\Gamma}_\mathrm{GD} + \mathbf{\Gamma}_\mathrm{Slon},
\label{eqn:LLG}
\end{eqnarray}
where $\mathbf{\Gamma}_\mathrm{LL}$ is the torque induced by an effective magnetic field which incorporates both the conservative and thermal stochastic components.  Gilbert damping $\mathbf{\Gamma}_\mathrm{GD}$ phenomenologically
describes the rate with which the energy is removed from the system.
Its torque is perpendicular to the Landau-Lifshitz one as well as to the SW orbits of the constant
energy.    The last term $\mathbf{\Gamma}_\mathrm{Slon}$
describes  the effect of the spin current  on the magnetization direction
\begin{eqnarray}
& &\mathbf{\Gamma}_\mathrm{LL} = -\gamma \mathbf{\hat{m}}\times \mathbf{H}_\mathrm{eff}, \nonumber \\
& &\mathbf{\Gamma}_\mathrm{GD} = -\gamma \alpha \mathbf{\hat{m}}\times\left[\mathbf{\hat{m}}\times \mathbf{H}_\mathrm{eff}\right], \nonumber \\
& &\mathbf{\Gamma}_\mathrm{Slon} = -\gamma  \mathbf{\hat{m}}\times\left[\mathbf{\hat{m}}\times \vec{\mathcal{I}}_s(t) \right].
\label{eqn:LLG_parts}
\end{eqnarray}
Here $\vec{\mathcal{I}}_s(t)=\mathbf{\hat{e}_p}\mathcal{I}_s(t)$ is the time dependent spin current in units of magnetization, polarized along the direction $\mathbf{\hat{e}_p}$ and $\gamma$ is the gyromagnetic ratio $\gamma = 176.1 \times 10^9 \mathrm{rad}/\mathrm{s}\, \mathrm{Tesla}$. For the DC perpendicular method $\mathcal{I}_s(t)$ is a rectangular pulse with duration $t_\mathrm{pl}$ and the amplitude $\mathcal{I}_s$, while for AC case we take it as $\mathcal{I}_s \sin(\omega t)$. An example of the AC spin current pulse is shown in Fig.~(\ref{fig:trajectory})a. For most simulations discussed here we took the pulse duration as three full periods $t_\mathrm{pl}=3(2\pi/\omega)$ of AC signal.
\begin{figure}[h]
  \begin{centering}
  \includegraphics[width=9cm]{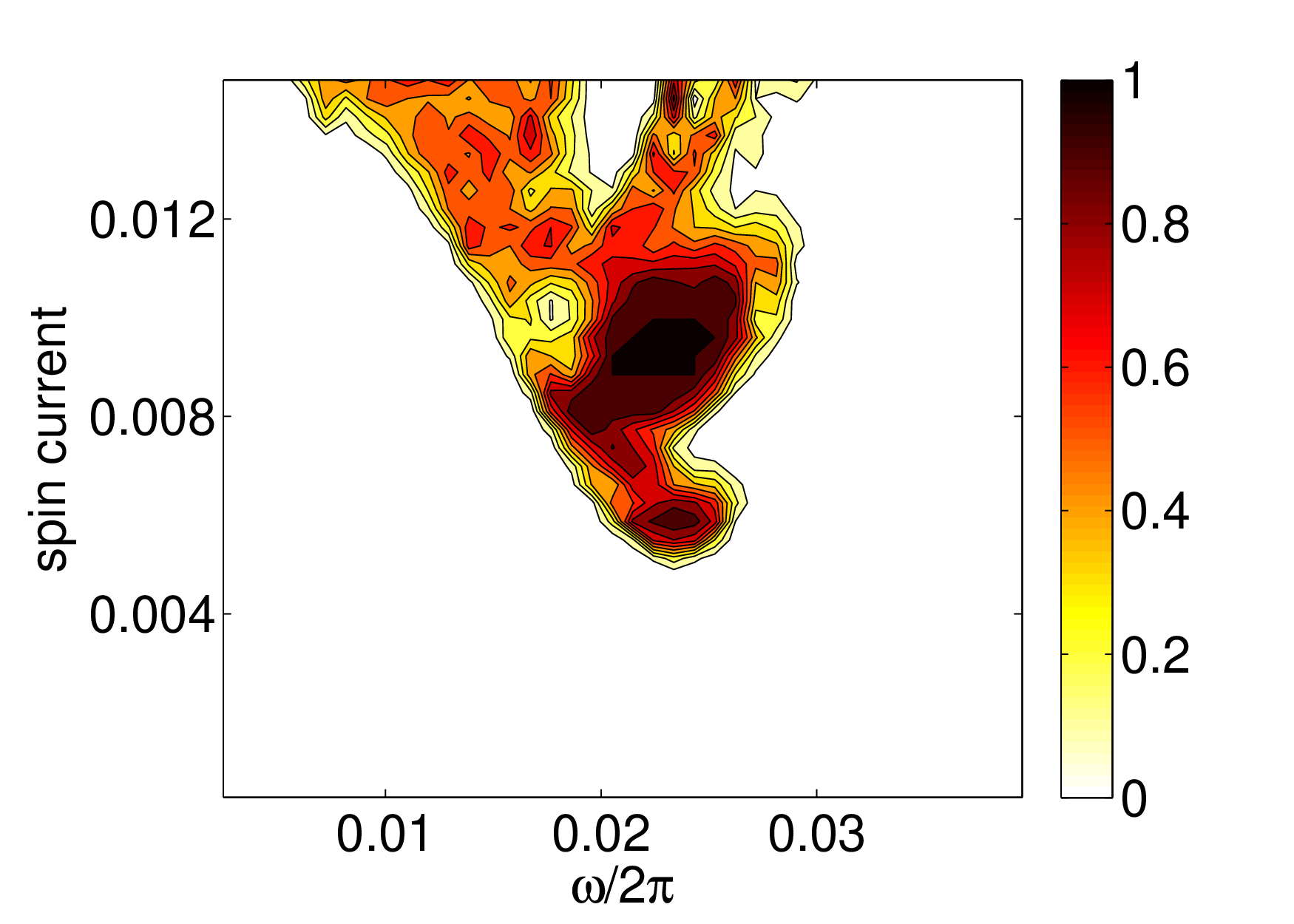}
  \par\end{centering}
  \caption{(Color online) Switching probability of AC perpendicular valve as a function of spin current amplitude ($\mathcal{I}_s/M_s$) and AC frequency $\omega/2\pi$. The pulse duration is taken to be three driving periods $t_\mathrm{pl} = 3(2\pi/\omega)$. Same parameters as in Fig.~\ref{fig:DC_probability}. }\label{fig:AC_probability}
\end{figure}

The effective magnetic field is given by
\begin{equation}
\label{eqn:H-eff}
\mathbf{H}_\mathrm{eff} (\mathbf{\hat{m}},t)= -\frac{1}{\mu_0 M_s}\, \nabla_{\mathbf{\hat{m}}} E(\mathbf{\hat{m}}) + \mathbf{h}(t)\,.
\end{equation}
Here $E(\mathbf{\hat{m}})$
is the conservative anisotropy energy, which we take as
\begin{eqnarray}
E(\mathbf{\hat{m}}) = \frac{ \mu_0 M_s}{2} \left( H_z  \left[1 - \left(\mathbf{\hat{m}} \cdot \mathbf{\hat{e}}_z \right)^2 \right]  +H_x \left( \mathbf{\hat{m}} \cdot \mathbf{\hat{e}}_x \right)^2 \right),
\label{eqn:energy}
\end{eqnarray}
where $H_z$ and $H_x$ are the anisotropy fields describing the easy axis and easy plane respectively.
The energy is zero $E=0$ along the easy axis directions $\pm \mathbf{\hat{e}}_z$ in the easy plane and reaches a
maximum $E = \mu_0 M_s(H_z+H_x)/2$ along the hard axis direction perpendicular to the easy plane $\pm\mathbf{\hat{e}}_x$. There
are two saddle points along the $\pm\mathbf{\hat{e}}_y$ axis, which separate the two minima and act as
the energy barrier with the energy $E_b =  \mu_0 M_s H_z/2$. The magnetization must overcome this barrier before the switching occurs.
Thermal noise is included via a random Gaussian magnetic field $\mathbf{h}(t)$ \cite{Brown63}  with the
correlator determined by the fluctuation-dissipation theorem \footnote{Non-equilibrium noise such as spin shot noise is omitted here as it is usually weaker than the thermal noise at room temperature \cite{Dunn10}}
\begin{eqnarray}
                                                      \label{eq:noise}
\left< \mathbf{h}_\mu(t) \mathbf{h}_\nu(t^\prime)\right> = 2\delta_{\mu\nu} \delta(t - t^\prime) \alpha T / \gamma M_s ,
\end{eqnarray}
where $T$ is the temperature of the free layer.


Using the above model we performed  numeric simulations of switching
processes for DC-perpendicular and AC-perpendicular spin currents.
We first present results for material parameters chosen as  $M_s =  1200 emu/cm^3$, $H_x=M_s$, $H_z = 0.033 M_s$,
$\alpha = 0.015$ and $T = 300 K$. Other choices of parameters are discussed after the theoretical part.
In both DC and AC perpendicular cases the spin current is polarized along the $z$ and $x$ axes
$\mathbf{\hat{e}}_p = \hat{\mathbf{x}} - \hat{\mathbf{z}}$ to simulate
the influence of the two fixed layers, Fig.~\ref{fig:MTJ}b. The $\mathbf{\hat{z}}$ component of the
spin current was chosen in the $-\mathbf{\hat{z}}$ direction to facilitate better switching behavior
in the DC case. This choice of z-axis polarization had little impact on the AC-perpendicular method switching
behavior.
For simplicity we assume equal magnitudes
for the perpendicular and parallel components the spin current. Their magnitudes may differ in practice
depending on the direction of the current through the spin valve and the materials used.

For the DC case Fig.~\ref{fig:DC_probability} shows the  switching probability as a function of spin
current amplitude $\mathcal{I}_s$ and the inverse pulse duration $1/t_\mathrm{pl}$.
One may notice that the switching current saturates at about $\mathcal{I}_s = 0.012 M_s$ in the limit of long pulses (the inverse period of small energy precessions is about $0.03 \gamma M_s)$. As was mentioned in the introduction the switch is most reliable then the pulse duration is less than half of the precession period. In this case the perpendicular torque is always acting in the proper direction of increasing the precession amplitude.
For  shorter pulses the critical current increases roughly as $t_\mathrm{pl}^{-1}$, the fidelity of the switch improves as well.

For the AC-perpendicular case Fig.~\ref{fig:AC_probability} shows the switching probability as a function of spin
current amplitude $\mathcal{I}_s$ and  AC frequency $\omega/2\pi$ with the pulse duration of three oscillation periods $t_\mathrm{pl}=3(2\pi/\omega)$, see also Fig.~\ref{fig:trajectory}a. Notice
that at optimal conditions the switching occurs at the current nearly two times lower than
that in the DC perpendicular case with the switching  currents as low as
$\mathcal{I}_s = 0.006M_s$. One should also notice the sharp frequency selectivity of the switching process. The optimal AC frequency appears to be somewhat below the natural small precession frequency $\Omega_0=(2\pi)\times 0.03 \gamma M_s$.

Simulations were also performed for pulses with different number of spin-current oscillations $t_\mathrm{pl} = n(2\pi/\omega)$.
For pulses with $n<3$ regions of low current switching were gradually excluded as $n$ approached $1/2$ until the switching probability was qualitatively the same as in the DC case
of Fig.~\ref{fig:DC_probability}. For pulses with $n>3$ the range of currents and frequencies which produced switching remained qualitatively the same as in Fig.~\ref{fig:AC_probability}. However, the switching probabilities are reduced and oscillate with increasing $n$ reaching probability of 50\% for sufficiently large values of $n$. This behavior is the result of magnetization switching back and forth between the two minima with longer pulse times.


\begin{figure}[h]
  \begin{centering}
  \includegraphics[width=9cm]{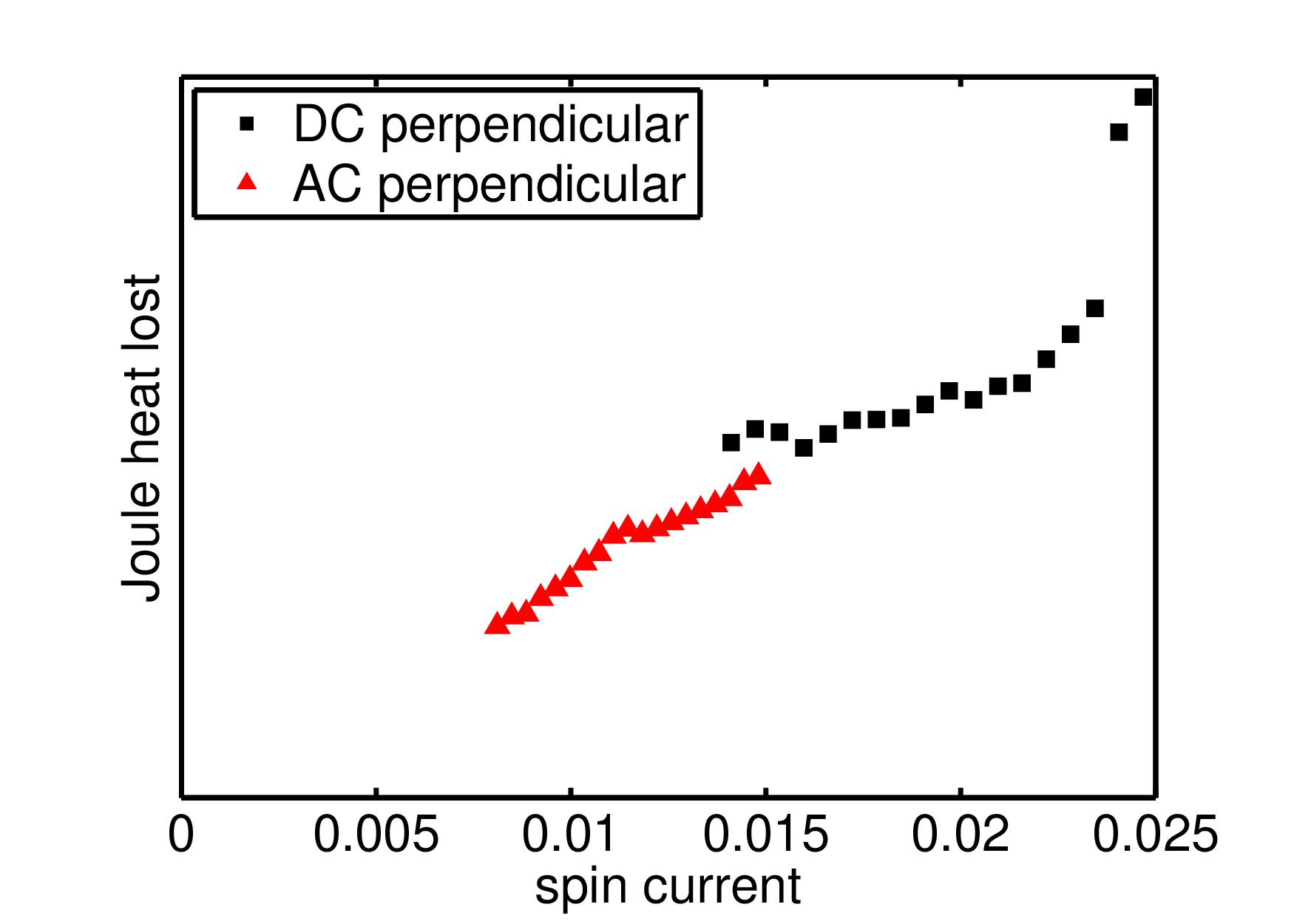}
  \par\end{centering}
  \caption{(Color online) Simulated optimal Joule heat lost during switching process in arbitrary units as a function of spin current amplitude ($\mathcal{I}_s/M_s$) for DC perpendicular (black squares) and AC perpendicular (red triangles) spin current methods. For AC simulations frequencies $\omega/2\pi$ ranged from $0-0.03 \gamma M_s$ with pulse times $t_\mathrm{pl} = n(2\pi/\omega)$ where $n$ ranged from $0.5-4$ to find the optimal regime. For both AC and DC cases only combinations of spin current $\mathcal{I}_s$ and AC frequency $\omega$ which result in switching probabilities above $90\%$ were are displayed. Same parameters as in Fig.~\ref{fig:DC_probability}.}\label{fig:Jouleheat_vs_current}
\end{figure}
Fig.~\ref{fig:Jouleheat_vs_current} shows the minimal Joule heat lost during the switch as a function of the spin current magnitude for both
the DC and AC perpendicular cases. It is calculated by numerically integrating the power over time
$\propto R\int_0^{t_{pl}} \mathcal{I}_s^2(t)dt$, where $R$ is a resistance which we assume to be constant through the switching process.
For both AC and DC cases only combinations of spin current $\mathcal{I}_s$ and AC frequency $\omega$ which result in switching probabilities
above $90\%$ were are displayed. Notice the AC perpendicular method cost significantly less energy to produce switching at low spin current
strengths. At its most energy efficient point the AC perpendicular method costs nearly one half the energy of the the most energy efficient
DC perpendicular method. This point occurs at driving frequency $\omega/2\pi \approxeq 0.024 \gamma M_s$, spin current amplitude
$\mathcal{I}_s \approxeq 0.007 M_s$, and the pulse duration of $t_\mathrm{pl} = 2(2\pi/\omega)$.

\section{Analytical considerations}
\label{sec:analytical}

\begin{figure}[h]
  \begin{centering}
  \includegraphics[width=9cm,angle=0]{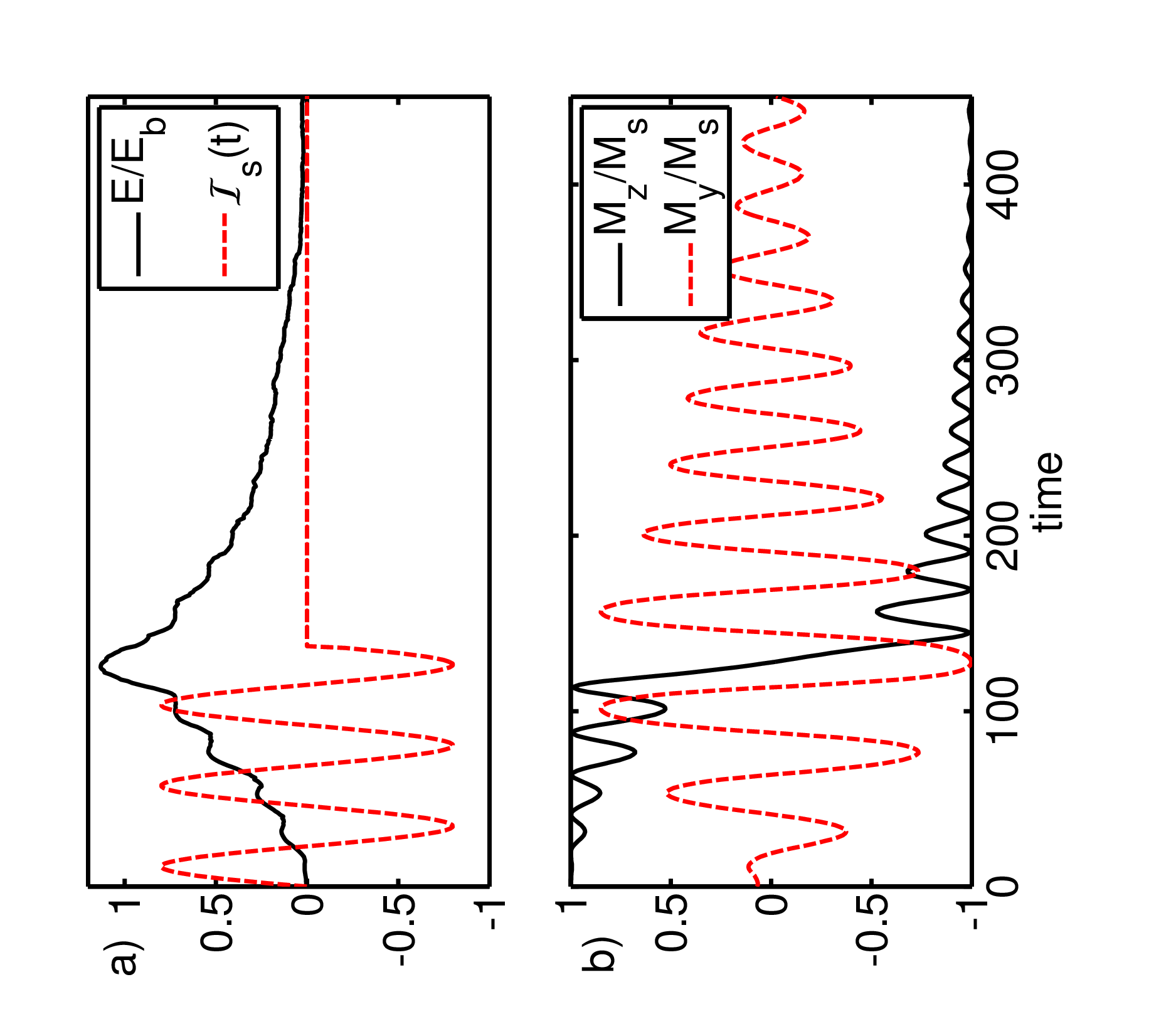}
  \par\end{centering}
  \caption{(Color online) a)  Spin current $\mathcal{I}_s(t)=\mathcal{I}_s\cos(\omega t)$ in arbitrary units (red, dashed) and normalized energy $E/E_b$ (black, solid) vs. time (in units of $(\gamma M_s)^{-1}$) for a successful switching event.  b) Magnetization $M_z/M_s$ (black, solid) and $M_x/M_s$ (red, dashed)  in units of $(\gamma M_s)^{-1}$ vs. time for the same event. Here $\mathcal{I}_s  = 0.005 M_s$ and $\omega/2\pi = 0.022\gamma M_s $. Other parameters are the same as in Fig.~\ref{fig:DC_probability}.}\label{fig:trajectory}
\end{figure}

We look now for a
qualitative understanding of why the AC protocol requires a significantly smaller critical spin current. Another feature
calling for an explanation is a sharp frequency dependence of the switching efficiency seen in Fig.~\ref{fig:AC_probability}.
Figure \ref{fig:trajectory}a shows the AC spin torque pulse $\mathcal{I}_s(t)$ and the time dependence
of the energy $E(t)=E({\bf \hat m}(t))$ along a switching trajectory. Notice that the energy increases  up to a bit over the top of the
energy barrier $E_b$ during the time of the AC pulse and later decreases down to zero in the state with the reversed magnetization.
Notice also that the energy $E(t)$ is a rather smooth function of time. On the other hand the magnetization ${\bf \hat m}(t)$,
depicted in Fig.~\ref{fig:trajectory}b, goes through three full revolutions before the switching, exhibiting highly non-monotonic behavior.

This observation suggests to parameterize
the magnetization vector $ {\bf \hat m}$ in terms of a slow variable $E$, which is nothing more
than the energy of the free layer given by equation \eqref{eqn:energy}, and a fast variable $\varphi$
which runs along a closed SW orbit of a constant energy.
It is  convenient to make the latter linearly related to the physical time (in the absence of the spin-torque).
To this end we identify $\varphi$ with the normalized time  required to reach a given point along  SW  orbit
within the purely conservative evolution, i.e. if $d {\bf \hat m}/dt = \mathbf{\Gamma}_\mathrm{LL}$ (we omit the
thermal stochastic noise (\ref{eq:noise}) throughout this section). With this choice one
takes $d\varphi =\Omega(E) dt$, where, cf.   Eq.~\eqref{eqn:LLG_parts},
\begin{eqnarray}
dt = \frac{\mathbf{\Gamma}_\mathrm{LL} \cdot d\mathbf{\hat{m}}}
{ \left|\mathbf{\Gamma}_\mathrm{LL}\right|^2} = \frac{d\varphi}{\Omega(E)}
\label{eqn:dt}
\end{eqnarray}
and $\Omega(E)$ is the frequency of SW orbit, given by $2\pi/\Omega(E)=\oint dt$,  where the integral runs along the SW
orbit with energy $E$ and $dt$ is given by Eq.~(\ref{eqn:dt}). This defines the {\em local} parametrization
${\bf\hat m}= {\bf\hat m}(E,\varphi)$, which may be shown \cite{Dunn12} to be implicitly specified by its partial derivatives:
\begin{equation}
                                                          \label{eq:derivatives}
\partial_\varphi {\bf\hat m} =  \frac{\mathbf{\Gamma}_\mathrm{LL}}{M_s\Omega(E)}  \,; \quad\quad
\partial_E {\bf\hat m} = \gamma \frac{\mathbf{\hat{m}}\times \mathbf{\Gamma}_\mathrm{LL}}
{ \left|\mathbf{\Gamma}_\mathrm{LL}\right|^2}\,.
\end{equation}
Notice that the parametrization is locally orthogonal, i.e. $\partial_\varphi {\bf\hat m}\cdot \partial_E {\bf\hat m} =0$.
One can thus project the identity $d {\bf \hat m}/dt = \partial_\varphi {\bf\hat m}\,\dot\varphi + \partial_E {\bf\hat m}\,\dot E$ onto these two
orthogonal directions and employ LLG equation~(\ref{eqn:LLG}) for $d {\bf \hat m}/dt$ to find equations of motion for $E$ and $\varphi$ pair
\begin{eqnarray}
\dot{E} &=& -\mathbf{H}_\mathrm{eff} \cdot \left(\mathbf{\Gamma}_\mathrm{GD} + \mathbf{\Gamma}_\mathrm{Slon} \right) \,;
\label{eqn:eone} \\
\dot{\varphi} &=& \Omega(E) - \Omega(E) \frac{\mathbf{\Gamma}_\mathrm{LL}\cdot \mathbf{\Gamma}_\mathrm{Slon}}{\left|\mathbf{\Gamma}_\mathrm{LL}\right|^2}\,.
\label{eqn:pone}
\end{eqnarray}
Here the right hand sides may be also expressed as functions of $E$ and $\varphi$. The energy evolution is driven only by relatively weak
damping and spin-torque terms, making the energy a slow variable. On the other hand, the phase dynamics is primarily given by a large $\Omega(E)$,
coming from LL term and making the phase the fast variable.

In presence of near resonant AC spin-torque $\mathcal{I}_s\sin (\omega t)$ the fast rotation phase $\varphi$ tends to follow the external drive,
see  Fig.~\ref{fig:trajectory}. One can thus introduce the relative phase $\phi(t)=\varphi(t) -\omega t$ as another slow variable.
The equation of motion for the slow phase $\phi(t)$ follows trivially from Eq.~(\ref{eqn:pone}) and the relation
\begin{eqnarray}
\dot{\phi} =  \dot{\varphi}-\omega
\label{eqn:deltadot}
\end{eqnarray}
One can now employ the fact that both $E$ and $\phi$  change relatively little through one precession cycle to
average over the fast precessing variable $\varphi$ \cite{Apalkov05,Dykman79}. To this end we write the AC spin current as
$\mathcal{I}_s\sin (\omega t)=\mathcal{I}_s\sin (\varphi-\phi)=\mathcal{I}_s(\sin\varphi \cos\phi - \cos\varphi\sin\phi)$ and integrate the
right hand sides of Eqs.~(\ref{eqn:eone}), (\ref{eqn:deltadot}) over $d\varphi$ with the help of Eq.~(\ref{eqn:dt}). This way we arrive at
the pair of equations of motion for the slow degrees of freedom
\begin{eqnarray}
\dot{E} &=& -\alpha U(E) - \mathcal{I}_s V(E)\sin\phi ,\nonumber\\
\dot{\phi} &=&  \Omega(E)-\omega  - \mathcal{I}_s  W(E) \cos\phi.
\label{eqn:dotave}
\end{eqnarray}
Here the generalized forces averaged over the orbit with energy $E$ are given by
\begin{eqnarray}
U(E)&=& \frac{\Omega(E)}{2\pi M_s}\oint\left[d\mathbf{M} \times \mathbf{H}_\mathrm{eff}\right]\cdot \mathbf{M} \,, \nonumber\\
V(E) &=& \frac{\Omega(E)}{2\pi M_s}\oint \left[d\mathbf{M}\times\mathbf{M}\right]\cdot \mathbf{\hat{e}}_p \cos \varphi\,,
                                                                \label{eqn:forces}                \\
W(E) &=& \gamma  \frac{\Omega(E)^2}{2\pi M_s}\oint \frac{ \left(d\mathbf{M}\cdot \mathbf{\Gamma}_\mathrm{LL}\right)
\left(\mathbf{\Gamma}_\mathrm{LL} \cdot \mathbf{\hat{e}}_p\right) }{|\mathbf{\Gamma}_\mathrm{LL}|^4}\sin \varphi, \nonumber
\end{eqnarray}
where in the last two expressions we used the parity  with respect to $\varphi$ to single out odd and even components.
It is worth noting here that the $\mathbf{\hat{z}}$ contribution to the spin current averages to zero upon this procedure.

\begin{figure}[h]
  \begin{centering}
  \includegraphics[width=9cm]{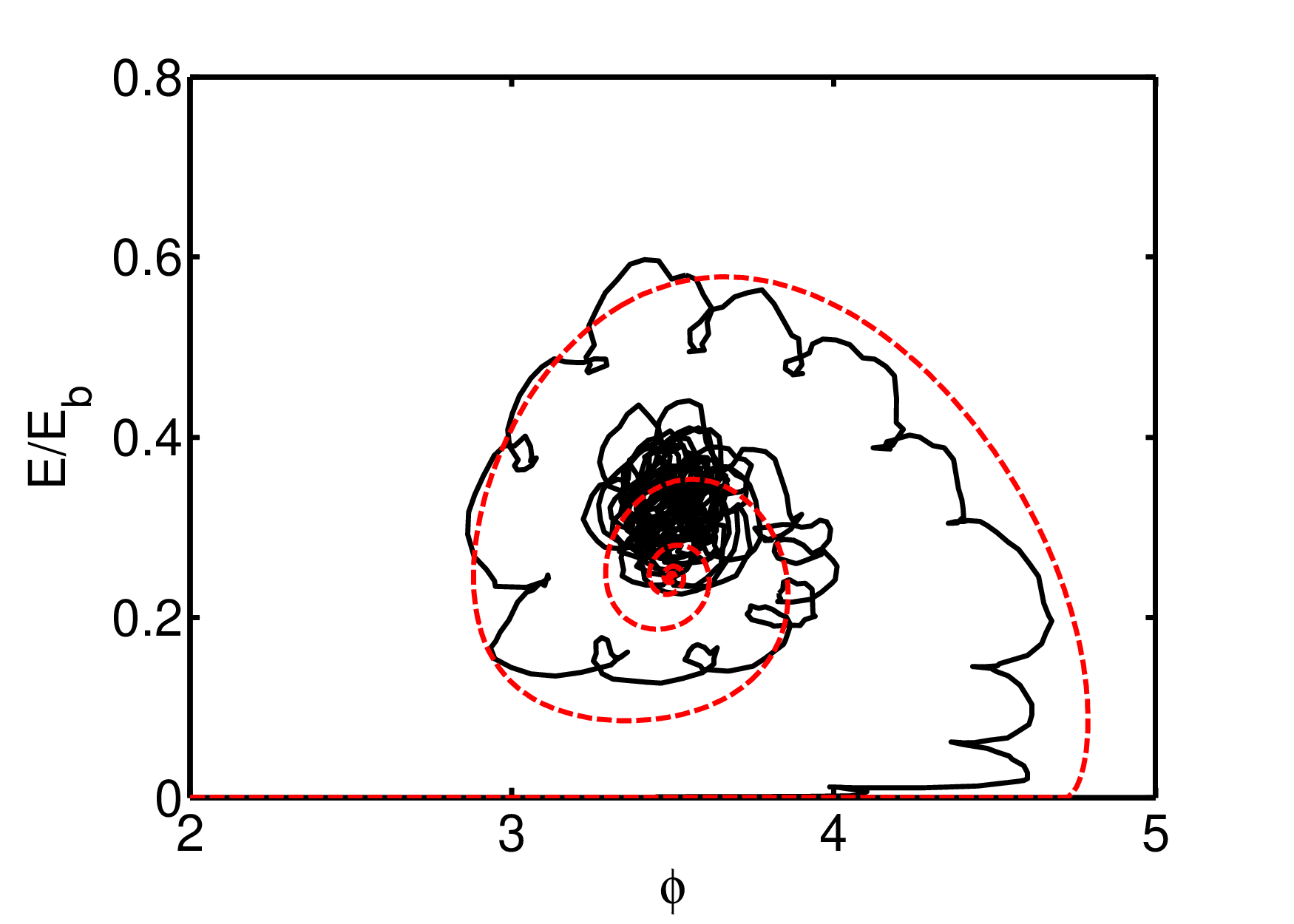}
  \par\end{centering}
  \caption{(Color online) Simulated energy $E$ vs phase $\phi$ (black, solid) calculated from numerical integration of the LLG equation \eqref{eqn:LLG} and average energy and phase (red, dashed) calculated from integration of equations \eqref{eqn:dotave}. Here $\mathcal{I}_s = 0.004 M_s$ and $\omega/2\pi = 0.028\gamma M_s$. Other parameters are the same as in Fig.~\ref{fig:DC_probability}.}\label{fig:ave_vs_real}
\end{figure}

A typical parametric trajectory  $E(t)$ versus $\phi(t)$ is shown in figure (\ref{fig:ave_vs_real}) for a subcritical spin current. Notice that after averaging over the fast fluctuations $E(t)$ and $\phi(t)$ behave similarly to coordinate/momentum pair  of an  underdamped oscillator. The oscillation frequency of this effective oscillator is much smaller than the precession frequency $\Omega(E)$ and is rather given by $\Omega(E)-\omega$. Of particular importance is the fact that before
winding down into the stationary point ($E_\mathrm{eq},\phi_\mathrm{eq}$), the energy highly {\em overshoots} its equilibrium value.
Had the energy exceeded the height of the energy barrier $E_b$ in the course of such an overshoot -- a switch is likely to occur.
This remains to be the case even if the equilibrium value $E_\mathrm{eq}$ is still below the barrier height $E_b$. This is
the qualitative reason for the AC critical current being significantly smaller than the DC one.

\begin{figure}
  \begin{centering}
  \includegraphics[width=9cm]{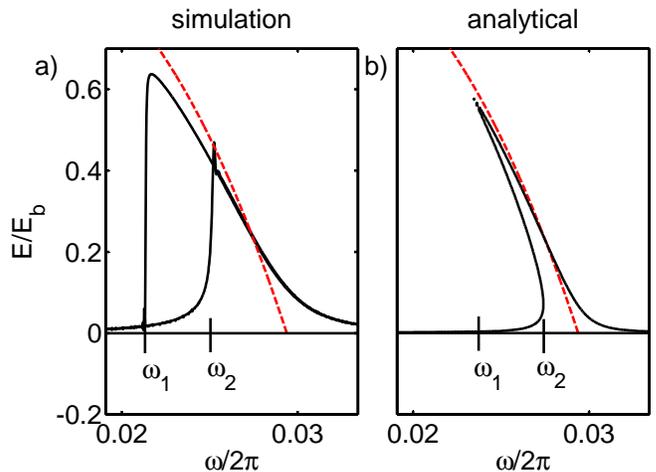}
  \par\end{centering}
  \caption{(Color online) Precession frequency $\Omega(E) = \omega$ (red, dashed) along with the equilibrium energy $E_\mathrm{eq}$ (black, solid) vs. driving frequency $\omega$ calculated from a) numeric integration of the LLG equation, Eq. \eqref{eqn:LLG} and b) Eqs. \eqref{eqn:equilibrium}. At frequencies  $\omega_1$ and $\omega_2$ the simulation jump between high and low energy driven precession modes. Here $\mathcal{I}_s = 0.002 M_s$ and $T = 0$. Other parameters same as in Fig.~\ref{fig:DC_probability}.}\label{fig:hyst}
\end{figure}

\begin{figure}
  \begin{centering}
  \includegraphics[width=9cm]{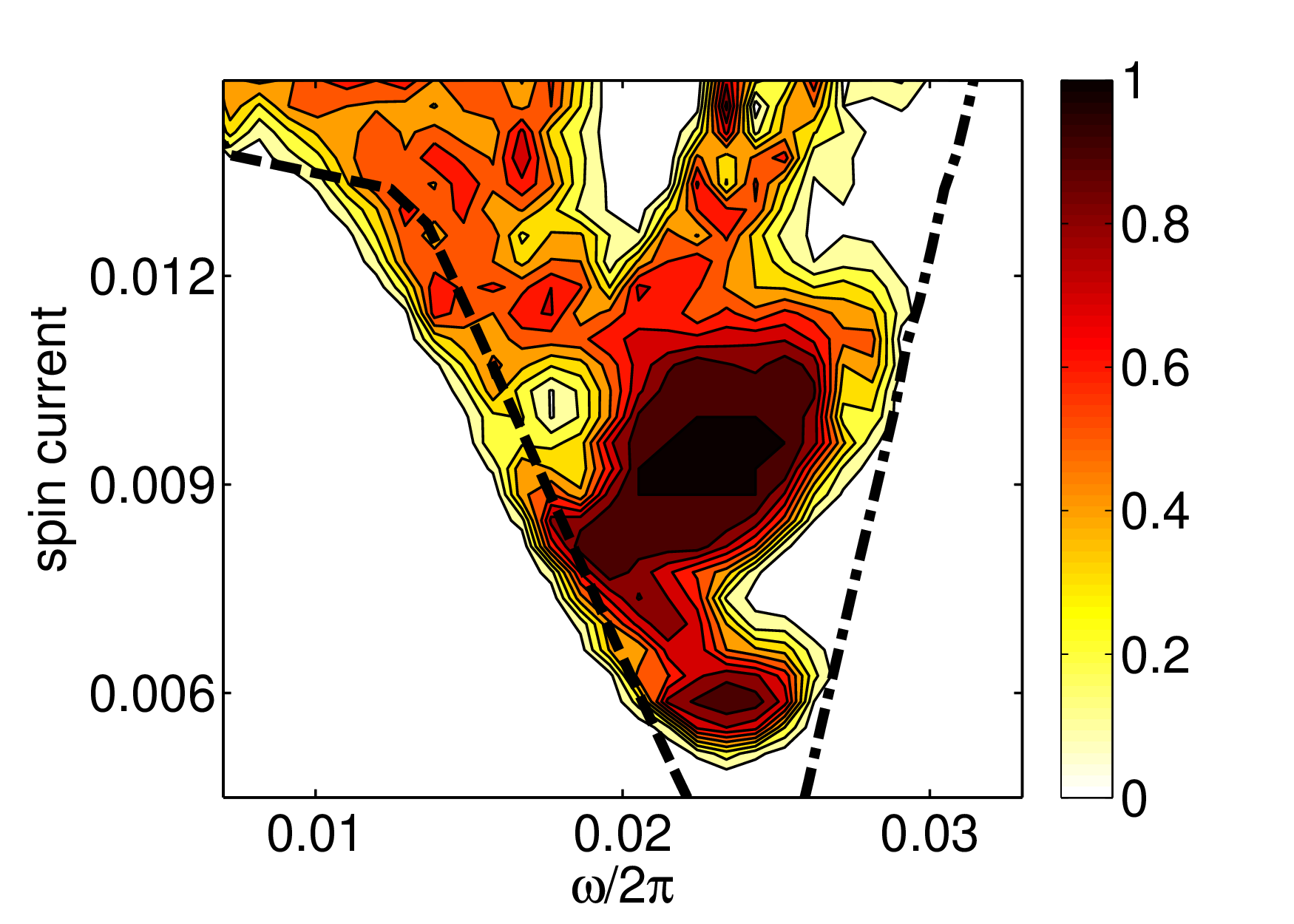}
  \par\end{centering}
  \caption{(Color online) Critical frequency $\omega =\omega_2$ (dashed) and critical current (dot-dashed) lines atop of the switching probability as a function of spin current ($\mathcal{I}_s/M_s$) and driving frequency (in units of $\gamma M_s$) for the pulse duration $t_\mathrm{pl}=3(2\pi/\omega)$.
  }\label{fig:boundries}
\end{figure}

To make progress in explaining sharp frequency dependence of the effect, one needs to look closely at the equilibrium point ($E_\mathrm{eq},\phi_\mathrm{eq}$)
of the effective oscillator. Such a point would be reached by the free layer magnetization upon continuous AC drive with the amplitude below the switching one after all transients decay due to dissipation.
Setting the left hand sides of equations \eqref{eqn:dotave} to zero gives such equilibrium conditions
\begin{eqnarray}
\mathcal{I}_s^2 &=& \left( \frac{\alpha U(E)}{V(E)} \right)^2 + \left( \frac{\omega - \Omega(E)}{W(E)}  \right)^2, \nonumber \\
\cot\phi &=& \frac{\alpha U(E) W(E)}{V(E)(\omega - \Omega(E))}.
\label{eqn:equilibrium}
\end{eqnarray}
Figure~\ref{fig:hyst} shows the equilibrium energy $E_\mathrm{eq}$ over a range of frequencies along with simulated result found by adiabatically varying $\omega$ over time. Notice that for the range of frequencies $\omega_1<\omega<\omega_2$ there are two stable solutions and an unstable branch in between. Consequently the simulation reveal hysteretic loop
with two {\it jump frequencies}, where the equilibrium energy abruptly jumps to a higher or lower value.
Since the goal is to reach the switching energy $E_b$,
one wants to insure that the oscillator follows the upper of the two branches, while having as large equilibrium energy as possible. As seen from
Fig.~\ref{fig:hyst} this is achieved at frequencies right above the upper hysteresis jump  $\omega_2\lesssim \omega$. We thus calculate the upper jump frequency of the hysteresis as a function of the spin current $\mathcal{I}_s$ and plot in Fig.~\ref{fig:boundries}.

To find a critical current of the swing switching
it is useful to notice that in the absence of the dissipation $\alpha\to 0$, the AC spin-torque driven oscillator~(\ref{eqn:dotave})
possesses an integral of motion. Indeed, one may check that the following function
\begin{equation}
                                                           \label{eq:Ham}
{\cal H}(E,\phi)=\int\limits_0^E dE'{\cal J}(E')\Big[\omega-\Omega(E')  + \mathcal{I}_s  \,W(E') \cos\phi \Big]  \,
\end{equation}
is conserved by the equations of motion~(\ref{eqn:dotave}), with $\alpha=0$, if the function $ {\cal J}(E)$ is a solution of the following
linear homogeneous differential equation:
\begin{equation}
                                                            \label{eq:jacobian}
V(E)\frac{d {\cal J}(E)}{dE} = {\cal J}(E)\left( W(E) - \frac{d V(E)}{dE}\right)\, .
\end{equation}
In fact one may define the action $I(E) =\int_0^E dE'{\cal J}(E')$, such that the change of
variables $(E,\phi)\to (I,\phi)$ in Eq.~(\ref{eq:Ham}) results in the Hamiltonian ${\cal H}(I,\phi)$ written in terms of the canonical action-angle pair. The
equations of motions~(\ref{eqn:dotave}) are nothing but the Hamilton equations:
$\dot I = \partial_\phi {\cal H}$ and $\dot \phi= -\partial_I {\cal H}$.
\begin{figure}
  \begin{centering}
  \includegraphics[width=9cm]{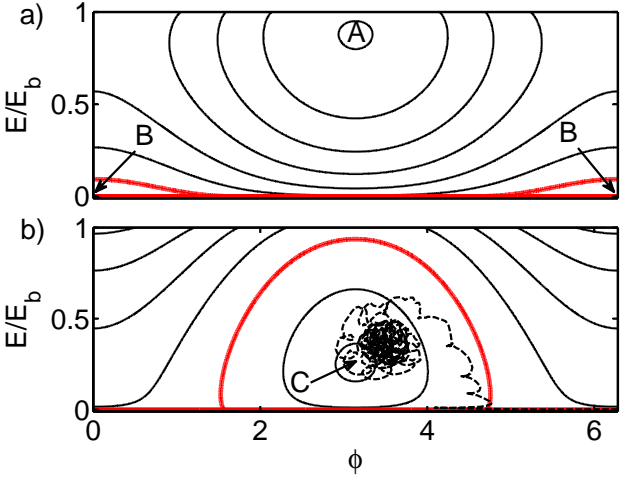}
  \par\end{centering}
  \caption{(Color online) Contours of a constant $\mathcal{H}$ for $\mathcal{I}_s = 0.003 M_s$ and for driving frequencies a) $\omega < \omega_2$ ($\omega/2\pi = 0.02\gamma M_s$) and b) $\omega > \omega_2$ ($\omega/2\pi = 0.028\gamma M_s$). Separatrix lines $\mathcal{H}=0$ are shown in bold red. Points A, B, and C correspond to equilibrium values  $E_{eq}$ and $\phi_{eq}$, cf. Fig.~\ref{fig:hyst}. Dashed black line in b) is a simulated trajectory for same parameters, already shown in
  Fig.~\ref{fig:ave_vs_real}.}\label{fig:contours}
\end{figure}

The consequence of this observation is that without the dissipation the $E,\phi$ trajectories are closed. The corresponding phase portraits are
are plotted in Fig.~\ref{fig:contours}a,b for the case of $\omega<\omega_2$ and $\omega>\omega_2$, correspondingly. There are two stable points $A$ and $B$ in the former case and only one $C$ in the latter. Hereafter we focus on the latter case, since for the former the evolution starting from $E=0$ can only bring the system to a small energy fixed point $B$ and thus is not beneficial for the switching. Starting from a vicinity of $E=0$ line the system evolves along a trajectory which is close to the separatrix line ${\cal H}=0$, before winding down towards the fixed point $C$ due to dissipation.  For the switching to take place one must require that the separatrix goes beyond the top of the barrier $E_b$. Since the maximum extent of the separatrix is reached at $\phi =\pi$, a necessary condition for switching to occur is:
\begin{equation}
                                          \label{eq:crit-current}
 \mathcal{I}_s > \frac{\int_0^{E_b}dE\, {\cal J}(E) (\omega-\Omega(E))}{ \int_0^{E_b}dE\,{\cal J}(E)W(E)}  \,.
\end{equation}
In fact this is an underestimate, since it does not take into account the dissipation. The latter dictates that the separatrix must exceed
$E_b+E_d$, where $E_d=\alpha \int dt U(E)=\alpha \int_0^{E_b}dE U(E)/\dot E$ is the work done by the dissipative force along the separatrix trajectory. One can now substitute $\dot E=- \mathcal{I}_s V(E)\sin\phi$, where $\sin \phi$ is found from the separatrix equation ${\cal H}=0$,
to find the dissipative loss $E_d$. One then uses $E_b+E_d$ instead of $E_b$ in Eq.~(\ref{eq:crit-current}) to obtain the minimal switching
current. The resulting critical line $\mathcal{I}_s(\omega)$ is plotted in Fig.~\ref{fig:boundries}. One can see that this line is indeed
close to the boundary of the observed switching range.

To make the general theory more transparent  it is useful to evaluate the averaged generalized forces (\ref{eqn:forces}) at small energy $E$. In this limits SW orbits are elliptical and the integrals in Eqs.~(\ref{eqn:forces}) may be easily performed, leading to
\begin{eqnarray}
                                                 \label{eq:small-energy}
\Omega(E) &=& \Omega_0\left(1 -  E/E_0\right)\,; \\
U(E)&=& uE\,;\quad V(E)=v\sqrt{E}\,;\quad W(E)=w/\sqrt{E} \,,\nonumber
\end{eqnarray}
where
\begin{eqnarray}
\Omega_0 &=& \gamma H_z \sqrt{1+h}, \quad\quad E_0=E_b(4+4h)/(2+h),\nonumber \\
u &=& \Omega_0 (2+h)/\sqrt{1+h} , \nonumber \quad \quad
v =\gamma \sqrt{E_b}\, \sqrt{1+h} , \\
w &=& \gamma \sqrt{E_b}\, \left(1+h-\sqrt{1+h}\right)/h . \nonumber
\end{eqnarray}
and $h=H_x/H_z$.
We have checked numerically that these expressions
remain almost quantitatively correct in the entire relevant range of parameters.
With these forces one finds ${\cal J}(E)=E^a$, where $a=(2w-v)/2v$ is eccentricity index which ranges from
$a=0$ in the absence of the easy plane anisotropy $h\to 0$ to  $a= -1/2$ in the strong easy plane case $h\gg 1$.

With these approximations the top of the separatrix line, given  by  the conditions ${\cal H}=0$ and $\phi=\pi$, is found from the cubic equation for $E^{1/2}$, cf. Eq.~(\ref{eq:Ham}),
\begin{equation}
                                                  \label{eq:separatrix}
\frac{a+1/2}{a+1}\, (\omega-\Omega_0)E^{1/2} +  \frac{a+1/2}{a+2}\, \frac{\Omega_0}{E_0} E^{3/2} = w\mathcal{I}_s.
\end{equation}
It may have either three or one solution with the bifurcation point determining the $\omega_2(\mathcal{I}_s)$ relation. The critical current
(\ref{eq:crit-current}) is then found by the substitution of $E=E_b$ in the left hand side of Eq.~(\ref{eq:separatrix}). This way one finds an
approximate location of the two lines depicted in Fig.~\ref{fig:boundries}. Their intersection results in the theoretical lower boundary
for swing-switching critical current, which happens to be practically $h$-independent for $h>1$
\begin{equation}
                                                  \label{eq:best-case}
\mathcal{I}_s \geq H_z/24\,.
\end{equation}
One may notice that in our simulations the critical switching current is about factor of $3$ larger than this idealized estimate. The reason is that
the estimate based on the conserved quantity (\ref{eq:Ham}) neglects the dissipation. Calculating the work of the friction force as described below Eq.~(\ref{eq:crit-current}), one finds $E_d\propto E_b\alpha (2+h)/ \sqrt{1+h}$ with the proportionality coefficient which logarithmically diverges at $\omega\to \omega_2$ (indeed, at the bifurcation point the dissipation always prevents the switching). In practice we found from the simulations that the condition
\begin{equation}
                                                  \label{eq:quality}
Q= \frac{1}{6\,\alpha} \, \frac{\sqrt{1+h}}{2+h} > 1\,
\end{equation}
provides a good estimate for the underdamped range ($E_d/E_b\lesssim 1$). If $Q\gg 1$ one expects to approach the theoretical boundary (\ref{eq:best-case}). In the opposite case $Q< 1$ the AC perpendicular method losses its advantage over the DC approach, because the effective oscillator is overdamped.

\begin{figure}
  \begin{centering}
  \includegraphics[width=9cm]{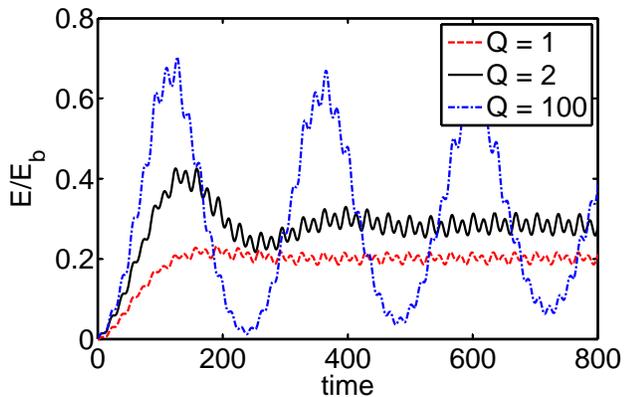}
  \par\end{centering}
  \caption{(Color online) Energy $E$ vs time (in units of $(\gamma M_s)^{-1}$) for $Q=100$ (blue, dot-dashed), $Q=2$ (black, solid), and $Q=1$ (red, dashed). Same parameters   as in Fig.\ref{fig:DC_probability} except of $\alpha$, which is changed to adjust $Q$.} \label{fig:damped}
\end{figure}

\begin{figure}
  \begin{centering}
  \includegraphics[width=9cm]{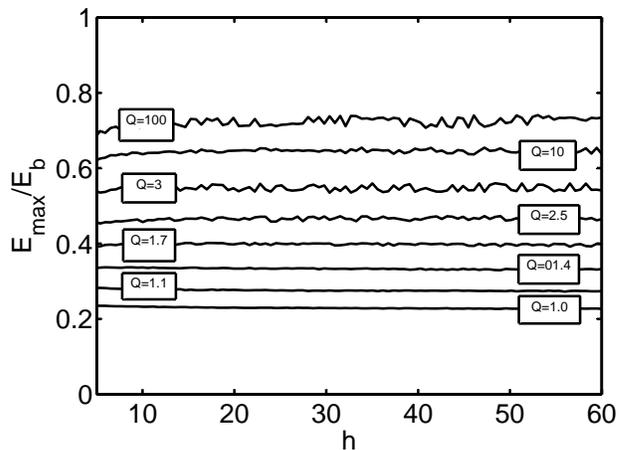}
  \par\end{centering}
  \caption{Maximum transient energy $E_{max}/E_b$ as a function of $h=H_x/H_z$ for several values of $Q$. Here $H_z = 0.033 M_s$, $\mathcal{I}_s = 0.003 M_s$, $T=0$, and $\omega=\Omega_0$.}\label{fig:evphi_compare}
\end{figure}

One may notice that the simulations discussed above happen to have a mediocre quality factor $Q\approx 2$, which may explain why they
result in the critical current which is  bigger than the theoretical bound (\ref{eq:best-case}). To test these predictions we performed additional simulations. Figure \ref{fig:damped} shows $E(t)$ for $Q=1$, $2$, and $100$. Notice that for $Q=1$ $E(t)$ is overdamped and for $Q=100$ $E(t)$ is highly underdamped. Notice also that in the underdamped $Q=100$ case the maximum energy $E_{max}$ is significantly higher than in the other two. Expanding this analysis further Fig.~\ref{fig:evphi_compare} shows the maximum transient energy $E_{max}$ as a function of $h$ for several values of $Q$.
To keep $Q$ constant we adjusted dissipation $\alpha$ according to Eq.~(\ref{eq:quality}), while $H_z$ and $\mathcal{I}_s$ are kept fixed and  $\omega=\Omega_0(h)$. Notice that the maximal energy remains almost completely constant for each value of $Q$ and grows with increasing $Q$, so that for $Q \gg 1$ the maximum energy is nearly $50\%$ larger than for $Q=2$. This shows that, if the spin-current is scaled with $H_z$ and AC frequency
is scaled with $\Omega_0$, the system behavior is governed by the {\em single} scaling parameter $Q$, defined in Eq.~(\ref{eq:quality}). This
observation reduces the multidimensional space of system's parameters down to the single essential parameter: the quality factor $Q$.


For cases where $h < 1$ the effectiveness of the AC-perpendicular method
is diminished in comparison to the equivalent DC method. This is
the result of higher energy trajectories, $E\simeq E_b$, spending more
time at larger values of azimuthal angle $\theta$. This causes the
$\mathbf{\hat{z}}$ contribution to the ST in the DC case to become
stronger than the $\mathbf{\hat{x}}$ contribution to the ST in the AC
case. A more detailed description of this effect is left for future works.

\section{Conclusion}

We have shown that by using an AC spin current with driving frequency close
to the precession frequency of the free layer one may achieve the reliable switching
of spin-torque valves. Keeping advantages of the DC perpendicular approach \cite{Lee09, Nikinov10, Liu10, Rowland11},
the switching times are well below $1ns$.  In addition, the AC method offers a significant reduction of the critical switching
current and associated energy dissipation. We explained the effect by the underdamped nature of the
effective non-linear oscillator, which allows to cross the energy barrier by transient oscillations.
The latter may significantly overshoot the equilibrium energy, reducing the spin-torque required for the switch to occur.

We have developed a theoretical description of the effect which allowed us to identify the optimal AC frequency
with the upper bifurcation frequency of the driven non-linear oscillator. It also provides with the estimate of the current amplitude needed to
perform the switch. Possibly most essentially, it provides the scaling arguments showing that  $\mathcal{I}_s\propto H_z$;
$\omega\approx \Omega_0=  \gamma H_z \sqrt{1+h}$, while the relative advantage of the AC method is governed by the single scaling
parameter: the quality factor $Q$, Eq.~(\ref{eq:quality}).

We would like to thank Alex Chudnovskiy for enlightening discussions.
This work was supported by NSF
Grant DMR-0804266 and U.S.-Israel Binational Science
Foundation Grant 2008075.


\end{document}